\documentstyle[11pt,iau185,twoside,epsf]{article}

\begin{document}
\title{Problems with the Pulsation Mode Selection Mechanism 
in the Lower
Instability Strip (Observations and Theory)}

\author{M. Breger\altaffilmark{1}, 
A.~A.~Pamyatnykh\altaffilmark{1,2,3}}
\affil{$^1$ Institut f\"ur Astronomie, Universit\"at Wien,
T\"urkenschanzstr. 17, A--1180 Wien, Austria; ~~email:
breger@astro.univie.ac.at} \affil{$^2$ Copernicus 
Astronomical Center, Bartycka
18, 00-716 Warsaw, Poland; ~~email: alosza@camk.edu.pl} 
\affil{$^3$ Institute of
Astronomy, Russian Academy of Sciences, Pyatnitskaya~48, 
109017~Moscow, Russia}

\begin{abstract}
We examine the severe disagreement between the number of 
predicted and observed
pulsation modes for $\delta$~Scuti stars. The selection of 
nonradial modes
trapped in the outer envelope is considered on the basis of 
kinetic energy
arguments. The trapped $\ell=1$ modes for the star 4~CVn 
are in good, but not
perfect agreement with the observations. The trapping of 
the $\ell=2$ modes is
weaker, so that this simple rule of mode selection may 
apply to $\ell=1$, and
possibly not to $\ell=2$ modes.

\end{abstract}

\keywords{Stars: oscillations, Stars: variables: 
$\delta$~Scuti stars}

\section{Introduction}

Delta Scuti star models predict pulsational instability in 
many radial and
nonradial modes. The observed number of low-degree modes is 
much lower than
the predicted number. The problem of mode selection is most 
severe for
post-main-sequence $\delta$~Scuti stars, which comprise 
about 40 percent of the
observed $\delta$~Scuti stars. The theoretical frequency 
spectrum of unstable
modes is very dense. Most modes are of mixed character: 
they behave like
p-modes in the envelope and like g-modes in the interior. 
For example, for a
model of 4 CVn we predict 554 unstable modes of $\ell$ = 0 
to 2: 6 for $\ell$ =
0, 168 for $\ell$ = 1, and 380 for $\ell$ = 2. However, 
only 18 (and an
additional 16 combination frequencies) were observed 
(Breger et al., 1999).
This is demonstrated in Fig.\,1. The problem for other 
$\delta$~Scuti stars,
such as BI~CMi, is similar.

\begin{figure}[h]
\plotone{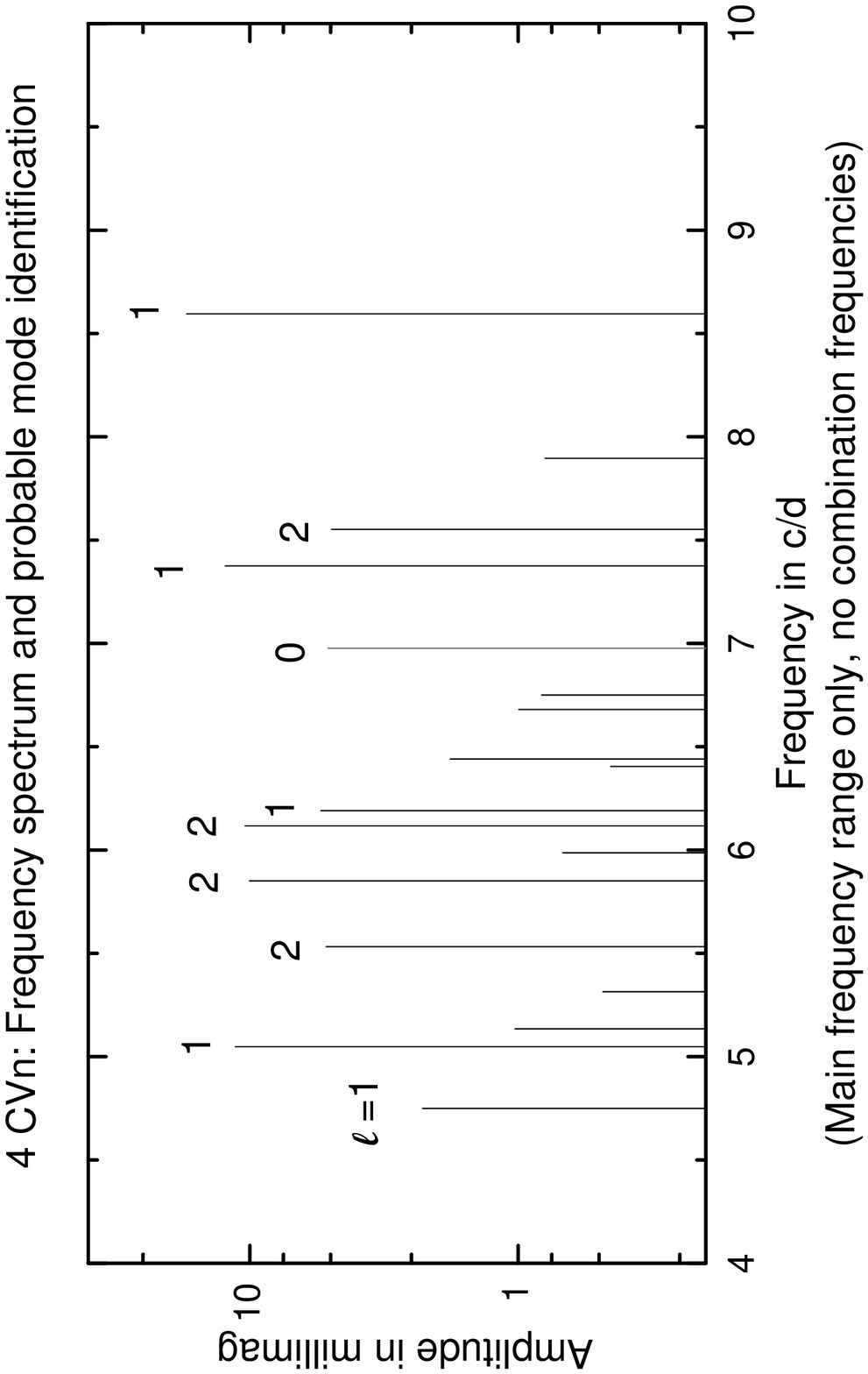}
\caption{Observed pulsation modes and probable mode 
identifications
for the evolved $\delta$ Scuti star 4~CVn.}
\end{figure}

Consequently, either the theoretical predictions or the 
observational
techniques are imperfect. On the observational side, only 
modes with
photometric amplitudes in excess of 0.5 mmag are observed. 
However, this does
not solve the dilemma, since the problem could then be 
rephrased to why only
some favored modes are excited to observable amplitudes, 
which usually are much
higher than the observational limit.

A number of different scenarios ranging from chaotic mode 
selection to energy
transfer between the modes through nonlinear mode coupling 
can be invoked to
explain the disagreement. Some of these hypotheses are very 
promising and are
amenable to observational tests:

\section{Scenario: random mode selection}

Here the selection of the star's observationally visible 
modes from the
multitudes of possible modes is random. A consequence of 
this scenario is
that we will not be able to make use of nonradial-mode 
frequencies for seismic
probing.

This hypotheses leads to several observational predictions: 
(i) there are no
regular frequency patterns of the photometrically visisble 
modes and, (ii)
there is no similarity of the frequencies of the visible 
modes from star to
star with similar masses, temperature and ages.

\section{Scenario: temporary growth of randomly selected 
modes to observable 
amplitudes by mode interaction}

In this scenario, most or all of the predicted modes are 
indeed unstable, but
with amplitudes too small to observe photometrically. 
However, due to mode
interaction, power is transferred between the modes. A few 
modes can
temporarily grow to high amplitudes and are observed. After 
several years,
which represent the typical growth cycle of $\delta$~Scuti 
stars, these modes
will decay again.

The hypothesis is supported by the fact that $\delta$ Scuti 
stars show
time-variable amplitudes as well as linear mode 
combinations of the modes with
high amplitudes. The amplitude variability may be so large 
that over a decade
the star may look like a completely different star (e.g. 
compare the power
spectra of 4~CVn during the 1966-1970, 1974-1978, 1983-1984 
and 1996-1997 time
periods). However, a detailed investigation of the data 
shows that modes do not
disappear to be replaced by others, but are still present 
at millimag
amplitudes. This argues against the hypothesis, at least on 
time scales of
decades.

\section{Scenario: trapped envelope modes}

In this scenario, the mode selection mechanism is related 
to modes trapped in
the envelope (Dziembowski \& Kr\'olikowska, 1990). It may 
be easier to excite
these modes with lower kinetic energy to a given amplitude 
on the surface. Such
a mode selection rule would be very simple and the 
computations rely only on
linear theory. However, the strong proof must come from the 
nonlinear theory.

\begin{figure}
\plotone{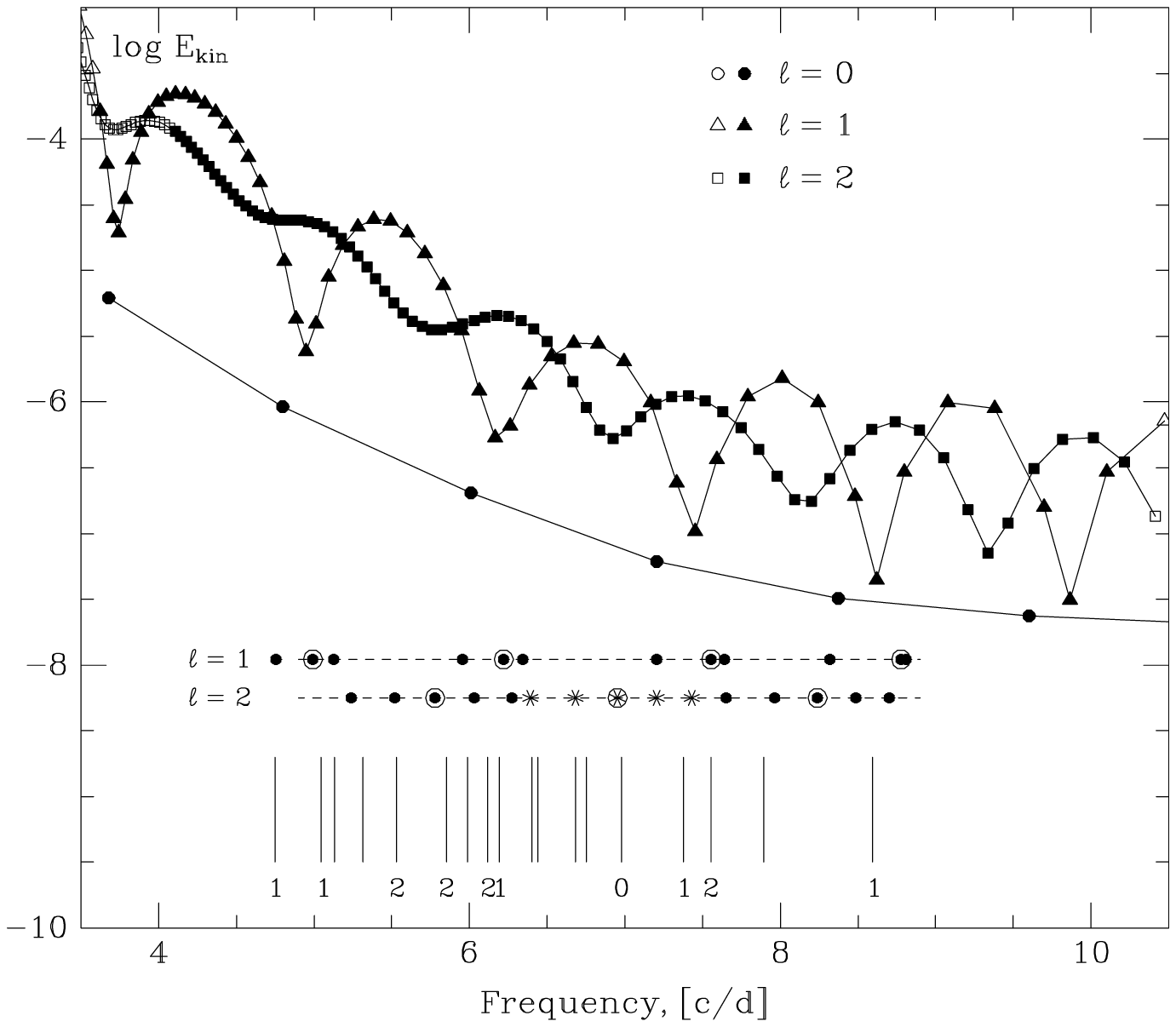}
\caption{Kinetic energy values associated with the unstable 
modes 
of the evolved $\delta$~Scuti star 4~CVn. 
The trapped modes are the modes with low kinetic energy 
values. 
Only the $m=0$ modes are shown. Empty symbols denote stable 
modes.
For the most trapped modes all components of the multiplets 
are shown 
separately and compared to the observed 
frequencies with probable mode identifications.}
\end{figure}

Fig.\,2 illustrates the complexity of the oscillation 
frequency spectrum for a
model of 4 CVn, which oscillates with at least 18 
frequencies in the range 4.7 -
9.7 c/d. The parameters of the model are within the range 
allowed by
observational data: 2.4 $M_\odot$, $\log L/L_\odot = 1.76$, 

$T_{\rm eff} = 6800K$, $\log g = 3.32$, $V_{\rm rot} = 82$ 
km/s. 
The chemical composition $X=0.70$, $Z=0.02$ was assumed. 
The ordinate gives
the oscillation kinetic energy, which is evaluated assuming 
the same radial
displacement at the surface for each mode. The high density 
of the oscillation
spectrum of nonradial modes is caused by very large values 
of the
Brunt-V\"ais\"al\"a frequency in the deep interior. This 
diagram is similar to
one given by Dziembowski (1997), with the observational 
mode identification and
selected rotational splitting added.

The most trapped modes are analogs of pure acoustic
modes. It can be seen that the trapping is much more 
effective for dipole
($\ell = 1$) than for quadrupole modes ($\ell= 2$). The 
trapping
effect for $\ell=2$ modes is especially weak at low 
frequencies and may not
apply here. Therefore, the selection rule might work for 
only $\ell=1$ modes
(see also Dziembowski \& Kr\'olikowska, 1990).

The observed spacing between identified $\ell=1$ modes is 
about 1.2 c/d,
corresponding to the theoretical spacing between trapped 
{$\ell=1$} modes.
Also, the rotational splitting for the trapped $\ell=1$ and 
also for $\ell=2$
modes is similar to the observed splitting for identified 
modes. Note,
however, that the conditions for the excitation of modes of 
different azimuthal
order $m$ may differ. Therefore, only some components of 
multiplets may be
excited to observed amplitudes.

For the model considered the theoretical frequencies of the 
best trapped modes
of $\ell=1$ fit the identified observed frequencies quite 
well. However, we
still consider this model to be only an illustrative one. 
To 
fit theoretical and
observed frequencies quantitatively, we must also take the
effect of the
rotational coupling of modes of different $\ell$ values 
into account.
Due to
this effect, close modes with the same $m$ and whose 
$\ell$ degrees differ by
2, are pushed away in frequency. This was demonstrated for 
a model of XX Pyx (Pamyatnykh et al. 1998).

In evolved $\delta$~Scuti models the trapping of nonradial 
modes in the
envelope can be more effective than in RR Lyrae models. For 
example, for the
most-trapped modes of $\ell=1$ in 4 CVn up to 67 percent of 
kinetic energy can be contributed by the envelope -- in 
comparison 
with less than 20 percent in the
RR Lyrae variables (see Dziembowski \& Cassisi, 1999).

\vspace{5mm}
It was shown that there is a high probability of a resonant 
excitation of
nonradial modes in radially pulsating RR Lyrae star models 
(see Nowakowski,
these proceedings and references therein). Very recently, 
nonradial modes were
detected in these stars (see Kov\'acs, these proceedings 
and references
therein). The theoretical frequency spectra of evolved 
$\delta$~Scuti stars are
similar to those of RR Lyrae stars, so resonances may be 
important for these
stars too.


\begin{references}
\reference Breger, M., Handler, G., Garrido, R., et al. 
1999, A\&A, 349, 225
\reference Dziembowski, W.A. 1997,
in Sounding solar and stellar interiors, 
eds.J.\ Provost \& F.-X.\ Schmieder, (Kluwer, Dordrecht), 
317
\reference Dziembowski, W.A. \& Cassisi, S. 1999, Acta 
Astron., 49, 371
\reference Dziembowski, W. \& Kr\'olikowska, M. 1990, Acta 
Astron., 40, 19
\reference Pamyatnykh, A.A., Dziembowski, W.A., Handler, 
G.,
\& Pikall, H. 1998, A\&A, 333, 141
\end{references}
\end{document}